\documentclass[12pt]{article}
\usepackage{amsmath,geometry,amsthm,amssymb,natbib,verbatim,upref,paralist,indentfirst,verbatim,graphicx,pdfpages,caption,hyperref}
\usepackage[T1]{fontenc}
\usepackage[utf8]{inputenc}
\usepackage{authblk}
\usepackage{tgheros}

\newtheorem{proposition}{Proposition}

\newtheorem{remark}{Remark}

\begin{document}
\title{Carbon Pricing and Resale in Emission Trading Systems}
\author{Peyman Khezr\thanks{School of Economics, Finance and Marketing, RMIT University, Melbourne, Australia. E-mail: peyman.khezr@rmit.edu.au.} }

\date{}

\maketitle

\begin{abstract}
Secondary markets and resale are integral components of all emission trading systems. Despite the justification for these secondary trades, such as unpredictable demand, they may encourage speculation and result in the misallocation of permits. In this paper, our aim is to underscore the importance of efficiency in the initial allocation mechanism and to explore how concerns leading to the establishment of secondary markets, such as uncertain demand, can be addressed through alternative means, such as frequent auctions. We demonstrate that the existence of a secondary market could lead to higher untruthful bids in the auction, further encouraging speculation and the accumulation of rent. Our results suggest that an inefficient initial allocation could enable speculators with no use value for the permits to bid in the auction and subsequently earn rents in secondary markets by trading these permits. Even if the secondary market operates efficiently, the resulting rent, which represents a potential loss of auction revenue, cannot be overlooked.
\end{abstract}

\vspace{1ex}
\noindent\textit{Keywords}: emission trading, carbon pricing, auction theory, market efficiency\\[1ex]
\noindent \textit{JEL Classification}: Q58, D44

\section{Introduction}

Emission trading systems (ETS) are critical tools in the global effort to combat climate change by providing a market-based approach to controlling greenhouse gas emissions. By setting a cap on total emissions and allowing firms to buy and sell emission permits, ETS create financial incentives for companies to reduce their carbon footprint. One prominent example is the European Union Emission Trading Scheme (EU ETS), which covers around 10,000 installations in various sectors across Europe. The EU ETS has been instrumental in reducing emissions by more than 30\% from 2005 levels, demonstrating its effectiveness. Another example is the Regional Greenhouse Gas Initiative (RGGI) in the United States, which includes several Northeastern states and has successfully reduced carbon dioxide emissions from the power sector by over 40\% since the beginning of the program in 2008. These examples illustrate how ETS can drive significant emission reductions, promote innovation in green technologies, and contribute to achieving international climate goals. 

Every emission trading system consists of frequent initial allocations of permits, usually through auctions, and a secondary market where firms can trade permits. Auctions play a crucial role in the initial allocation of emission permits within emission trading systems, even in the presence of secondary markets. The primary advantage of auctions is their ability to allocate permits efficiently to firms that value them the most, thereby promoting cost-effective emissions reduction \citep{Cramton2002}. By setting a transparent price discovery mechanism, auctions ensure that permits are distributed based on firms' willingness to pay, reflecting their marginal abatement costs \citep{Cramton2002, tietenberg2010emissions}. This market-driven approach helps avoid the pitfalls of free allocations, such as windfall profits and perverse incentives for high emitters.

Moreover, auctions generate significant revenue that can be used to fund environmental programs or mitigate the economic impacts of transitioning to a low-carbon economy. This contrasts with secondary markets, which, while providing flexibility and liquidity, do not directly generate such revenues for regulatory bodies. Also, auctions, if designed properly, help prevent market manipulation and speculative behavior that might distort prices in secondary markets, ensuring a more stable and predictable permit pricing system.

In this paper, we investigate the role of secondary markets in ETS and how they influence bidding behavior in the auction. We first show that secondary markets could contribute to a larger level of demand reduction in the auction, consequently reducing the auction efficiency and increasing rents. Second, we show that the efficient assignment of permits in the auction must be the main goal, which eventually results in the highest social surplus. This would minimize the rent by not allowing speculation in the auction. Finally, we show that if such an assignment exists, the role of secondary markets could be easily addressed by running frequent auctions and allowing reasonable banking of permits.

\section{Background and Previous Literature}

Emission Trading Systems have been studied extensively as mechanisms for reducing greenhouse gas emissions in a cost-effective manner. The theoretical foundation of ETS is based on the premise that setting a cap on emissions and allowing trading can lead to efficient pollution control \citep{Montgomery1972}. Montgomery's work laid the groundwork for understanding how permits could be traded in a market setting, providing firms with the flexibility to meet regulatory requirements in a cost-minimizing fashion.

Subsequent research has focused on the design of the auctions used to distribute these permits. Auctions are favored for initial allocation due to their ability to discover prices transparently and allocate permits efficiently \citep{Cramton2002, Khezr2018a}. For instance, \cite{Cramton2002} studies a clock auction format and the implications for ETS, emphasizing the importance of auction design in achieving economic and environmental goals.

The most commonly adopted auction in cap-and-trade markets is the uniform-price auction, which is favored for its simple rules and effective price discovery properties \citep{Khezr2018b}. Despite its popularity, it is widely known that such auctions do not promote truthful bidding; participants often bid below their actual valuation to manipulate market outcomes, a phenomenon known as demand reduction \citep{Back1993, Ausubel2014}. One possible way to address these inefficiencies is through a secondary market. A properly functioning secondary market ensures that permits are eventually allocated to those who value them the most. However, during the trading process, there will be a transfer of rents among various parties. Therefore, despite the existence of secondary markets, the efficiency of the initial allocation remains as a crucial market design question.

There are several studies that use various designs to address the inefficiency of the uniform price auction \citep{Back2001, McAdams2007, khezr2021revenue}. These designs would mainly require an adjustment of the supply of permits to a type of supply that is not perfectly inelastic.\footnote{There are few examples where the design is via an slight adjustment of the payment rule rather than the supply. For example see \cite{khezr2020semi}. } For instance, \cite{McAdams2007} argues that flexibility in the supply quantity announced prior to the auction could help reduce incentives for demand reduction, thus promoting more honest bidding behaviors. In real world emissions markets these increasing supply schedules are implemented using price caps. For instance \cite{friesen2020mind} conduct lab experiments to study the use of price caps for the case of RGGI.\footnote{Various studies use experiments to study uniform-price auctions and how they perform in cap-and-trade markets \citep{shobe2010experimental, Shobe2014402, Holt2016, perkis2016experimental}.}

The interaction between primary auction markets and secondary trading has also been a significant area of research. Secondary markets are essential for providing liquidity and allowing for adjustments as firms’ needs change over time \citep{Hahn1984}. Hahn's analysis highlights how secondary markets facilitate the efficient redistribution of permits, enhancing the flexibility and cost-effectiveness of markets such as ETS.

\section{Model}

There are $n\geq 2$ firms active in a market for emissions that are potential buyers of emission permits. There are $k$ homogeneous number of permits available for sale via a uniform-price auction. Suppose two types of firms are active in the market: polluters and speculators. Polluters are those that demand permits for their own use and speculators are those that only demand permits for speculation purposes to make a profit from dealing permits. For simplicity assume there is only one speculator firm denoted by firm $s$. The identity of firms are their own private information. So they cannot distinguish the speculator. The value of $k$ permits to each firm is known privately. For simplicity suppose each firm $i$ has a constant private $v_i$ for all the $k$ units. Each $v_i$ is i.i.d according to a commonly known distribution function $F(.)$. The speculator has zero use value for the permits in the auction and only knows the distribution of values for other firms. 

The market is divided into two main branches: an auction and a secondary market. In the auction, the regulator first allocates $k$ number of permits via a sealed-bid uniform-price auction. In the secondary market, firms engage in mutual trades to trade permits at the market price.

\subsection{An illustrative example}

Suppose there are four bidders each with two units of demand and there are total four units available for sale. Suppose the values of each bidder is as shown in Table \ref{tab1}. First, if all bidders bid truthfully the auction clearing price becomes 6 with a revenue equal to 24. Also the bidders' surplus in this case is equal to 10. Note that in this case a speculator cannot make a positive profit by bidding in the auction. To see this, look at the marginal bidder who loses an item due to a bid by the speculator. Since that bidder would now set the price, and since the bid is equal to the value of the bidder, it is not possible for the speculator to make a positive profit by reselling that unit the the bidder: any positive profit for the speculator requires a trade at a price above the bidder's value.

\begin{table}[h!]
\caption{Example 1}
\label{tab1}
\begin{center}
\begin{tabular}{|l|c|c|c|c|c|c|c|c|c|}
\hline
&$v/b$& $B_1$ & $B_2$ & $B_3$ & $B_4$& s &$p^*$& Surp.&Rev. \\ \hline
True values&$(v_1, v_2)$ & $(10,9)$ & $(8,7)$ & $(6,5)$ & $(4,3)$ &$(0,0)$&6&&\\ \hline
&payoffs &  7& 3 & 0 & 0&&&\textbf{10} & 24\\ \hline
e.g. bids &$(b_1, b_2)$ & $(10,7)$ & $(8,5)$ & $(6,4)$ & $(4,2)$&$(0,0)$&5&& \\ \hline
&payoffs &9  & 3 & 1 & 0&&&\textbf{13}&20\\ \hline
bids with $s$ &$(b_1, b_2)$ & $(10,7)$ & $(8,5)$ & $(6,4)$ & $(4,2)$&$(6.5,0)$&6&&\\ \hline

&payoffs &7  & 2 & 0 & 0&&&\textbf{9}& 24 \\ \hline
\multicolumn{9}{|l}{Higher demand reduction} & \\\hline
e.g. bids &$(b'_1, b'_2)$ & $(10,5)$ & $(8,4)$ & $(6,3)$ & $(4,2)$&$(0,0)$&4&& \\ \hline
&payoffs &11  & 4 & 2 & 0&&& \textbf{17}&16\\ \hline
bids with $s$&$(b'_1, b'_2)$ & $(10,5)$ & $(8,4)$  & $(6,3)$ & $(4,2)$&$(6,0)$&5&& \\ \hline
&payoffs &5  & 3 & 1 & 0&&& \textbf{9}& 20\\ \hline
\end{tabular}%
\end{center}
\end{table}

Next consider the second case where bidders reduce their demands by following a bidding strategy $(b_1, b_2)$ (Table \ref{tab1}). The auction price declines to 5 and the revenue reduces to 20. However, the surplus increases to 13 and less than the decrease in revenue. The reduction in overall surplus is due to inefficient allocation of units. Now if a speculator enters the auction and submits a bid equal to 6.5 for only one unit, then the auction outcome changes as follows. The price becomes 6 with a revenue equal to 24, similar to the truthful case. However, bidders' surplus declines by one unit. Note that at this stage there is no surplus for the speculator because she places zero value for the item. Given that the speculator has an item purchased at a price equal to 6 and player 2 did not won a second item while she has a value of 7 for that, there is a possibility of a trade in the secondary market. It is straightforward to check that the sum of surplus of player 2 and the speculator for this trade is equal to 1. Thus the outcomes of the auction with demand reduction and the secondary market becomes similar to the truthful situation. However, in any trade the speculator receives a positive surplus and player 2's surplus becomes strictly lower than the truthful allocation.

Consider a third case where bidders reduce their demand even further by following the bidding strategy $(b'_1, b'_2)$ (Table \ref{tab1}). Without a speculator, the price declines more than the previous case to 4. As a results the surplus increases to 16 and the auction revenue also becomes 16. Therefore the overall surplus declines even further due to more inefficient allocation of units. The speculator can win a unit by submitting a bid above 5. However, the auction clearing price raises to 5 and bidders' surplus becomes 9. In this case the overall surplus declines to 29. While the possibility of a trade in a secondary market can raise the surplus by another 4 units ---value of player 1 for the second unit is 9 and the value of the speculator is 5--- there will be a transfer of surplus from player 1 to the speculator. 

As shown in this example, the sum of revenue and bidders' surplus is the highest when we have truthful bidding. With a secondary market --- either with or without the regulator--- it is possible to achieve the efficient allocation if the initial allocation via the auction in not efficient due to demand reduction. However, there are still two main issues even if a secondary market achieves efficiency. First, the regulator's revenue certainly declines if there's no speculator. Second, a speculator `may' increase the revenue back to the truthful bidding but there will be a positive transfer of surplus to the speculator who has zero value for the units. 

\section{Results}

First consider a case without a secondary market where all the permits are allocated via the auction and firms do not have a second chance to trade permits. It is very well-known that in any undominated equilibria of the uniform-price auction, firms submit untruthful demands that are below their actual demand \citep[][Ch. 13]{Krishna2002}. Denote $\mathbf{b}=(\mathbf{b}_1,..., \mathbf{b}_n)$ as a representative equilibrium bids (submitted demands) for each firms. In particular $\mathbf{b}_i$ is the submitted demand schedule by bidder $i$ in equilibrium. Further denote $p^*$ as the equilibrium clearing price given the submitted demands and total quantity of permits $k$. The following proposition explores a situation where there is a secondary market. 

\begin{proposition}
Given any undominated equilibrium of the uniform-price auction $\mathbf{b}$, in any new equilibrium with a secondary market $\mathbf{b}'$, firms submit weakly lower demand schedules and the new equilibrium price is $p' \leq p^*$.
\end{proposition}

\begin{proof}
Start backward from the secondary market. Suppose firm $i$ with a value $v_i$ wants to purchase a permit in the secondary market from firm $j$ who holds a permit and values it $v_j<v_i$. This is essential for any possible trade. Suppose the trade takes place at a price $v_j<p<v_i$. Thus firm $i$'s payoff from this trade is equal to $v_i-p$. Denote $b_i$ as firm $i$'s equilibrium bid for the same unit in the auction without a secondary market. It is enough to show in any new equilibrium the new bid $b'_i$ is less or equal than $b_i$. Further denote $p^*$ as the equilibrium price for $\mathbf{b}$. Consider firm $j$ with a value vector $\mathbf{v}^j = (v_1, v_2,.., v_j,..,v_k)$ who wins $j-1$ units. If $v_j<p^*$ then the payoff of winning the $j$th unit is negative. However, if $v_j>p^*$ then firm $i$ could attain a positive payoff by winning the $j$th unit. 

Secondary Market Dynamics: Assume that $v_j > p^*$ allows firm $j$ to win the auction. After the auction, firm $j$ might be willing to sell the permit in the secondary market if it can make a profit. For firm $i$, who values the permit more than the auction clearing price but less than $v_j$, there is an incentive to buy this permit if it is offered at a price $p'$ such that $v_i > p'$. This transaction is beneficial for both $i$ and $j$ if $v_j < p' < v_i$. 

Effect on Bidding Strategies: Knowing that there is an opportunity to buy permits on the secondary market can influence firm $i$'s initial bidding strategy in the auction. Since firm $i$ anticipates the possibility of purchasing permits at a price potentially lower than $v_i$ (due to other firms’ willingness to sell for profits), it might bid less aggressively in the primary auction. This strategic underbidding is rational as it allows firm $i$ to minimize expenditure while still securing the necessary permits either in the auction or from the secondary market.

Lower Bids Lead to Lower Clearing Prices: As more firms adopt this underbidding strategy, the overall demand reflected in the auction decreases, which may lead to a lower clearing price $p' \leq p^*$. Essentially, the presence of the secondary market introduces additional flexibility and potential for arbitrage, prompting firms to adjust their bids downwards in the primary auction to leverage cheaper buying opportunities post-auction.

Generalization of the Effect: This underbidding behavior, when generalized across many firms participating in the auction, leads to a systematic reduction in the equilibrium bids $\mathbf{b}'$, where each component of $\mathbf{b}'$ is weakly lower than the corresponding component in $\mathbf{b}$. Consequently, the new equilibrium price in the auction, denoted $p'$, is lower than or equal to the original clearing price $p^*$, reflecting the reduced willingness to pay in the auction due to the option to trade later.

Thus, given any undominated equilibrium of the uniform-price auction $\mathbf{b}$, in any new equilibrium with a secondary market $\mathbf{b}'$, firms submit weakly lower demand schedules and the new equilibrium price $p'$ is less than or equal to $p^*$.
\end{proof}

The above proposition suggests that a secondary market could contribute to the extent of demand reduction. While there are no strict connections between the extent of demand reduction and inefficiency of the allocation, on average, higher demand reduction means lower revenue and lower clearing prices. Also higher demand reduction increases the chance of inefficient allocation of permits given that buyers with lower values would have a greater chance to submit a successful bid. This incident will become more important when bidders do not reduce their demands monotonically.

\begin{proposition}\label{pro2}
A multi-unit assignment $\varphi$ with a payment rule $\Phi$ would maximize the total surplus if and only if it is efficient. 
\end{proposition}

\begin{proof}
We define total surplus as the sum of consumer surplus and producer surplus. Let's denote the valuation of firm $i$ for a unit of permit as $v_i$ and the payment it makes under the rule $\Phi$ as $\Phi_i$. The total surplus (TS) is then given by:

\[
TS = \sum_{i} (v_i - \Phi_i) + \sum_{i} \Phi_i = \sum_{i} v_i
\]

This formulation shows that the total surplus is simply the sum of the valuations of all firms that receive the permits, irrespective of the payment transferred to the auctioneer, as payments are transfers and do not affect the total surplus.

\textbf{Efficiency}: We say that an allocation $\varphi$ is efficient if it allocates permits to the $k$ highest valuers among the firms. Let $\varphi^*$ be an efficient allocation. For any other allocation $\varphi'$ that is not efficient, there exists at least one firm $j$ such that $v_j > v_i$ where firm $i$ receives a permit under $\varphi'$ but not under $\varphi^*$, and $j$ does not receive it under $\varphi'$ but does under $\varphi^*$. Then:

\[
\sum_{i \in \varphi'} v_i < \sum_{i \in \varphi^*} v_i
\]

\textbf{Necessity}: If an allocation $\varphi$ maximizes the total surplus, then it must allocate permits to the firms with the highest valuations. Assume for contradiction that $\varphi$ does not allocate to one of the highest valuers, say $j$, and instead allocates to a lower valuer, $i$. If we reallocate the permit from $i$ to $j$, the total surplus increases by $v_j - v_i$, which is positive by assumption. Hence, a surplus-maximizing allocation must be efficient.

\textbf{Sufficiency}: Next, assume $\varphi$ is efficient. By definition, it allocates permits to the firms with the highest valuations. Therefore, no reallocation can increase the total surplus, as any reallocation would involve moving a permit from a higher valuer to a lower one, which would decrease the total surplus.

Therefore, $\varphi$ maximizes the total surplus if and only if it is efficient, as reallocating permits within an efficient allocation does not increase the total surplus and any deviation from an efficient allocation reduces the total surplus.
\end{proof}

Proposition \ref{pro2} highlights the importance of the initial allocation in achieving highest total surplus. In fact, only a fully efficient assignment can achieve this goal. Now the question is, what if the initial allocation is not efficient but the secondary market is? The next proposition is going to investigate such cases.

\begin{proposition}
 When the initial assignment $\varphi$ is not efficient, an efficient secondary trade would maximize the total surplus with strictly positive rents.
\end{proposition}

\begin{proof}
    Denote $\varphi'$ as the efficient allocation of the secondary market. Because $\varphi$ was inefficient, there exist a buyer $i$ with a value $v_i$ who did not receive a unit, and a buyer $j$ with a value $v_j< v_i$ who receives a unit. Suppose this is the only instance of inefficiency. In the secondary market and assignment $\varphi'$, it must be the case that $i$ is allocated the unit. For any possible trade in the secondary market between $i$ and $j$, there must exist a price $v_j< p <v_i$, otherwise either $j$ would be better off by keeping the unit or $i$ would be worst off by obtaining the unit. After the transfer between $i$ and $j$ at a price equal to $p$, $j$'s receives a payoff equal to $p-v_j$. This amount is equal to the rent that $j$ receives due to an inefficient initial allocation.   
\end{proof}

While a positive rent may seem a secondary concern when the total surplus is maximized, it can have significant consequences on the market. For instance, if thousands of permits are misallocated in the initial allocation, the rent in the secondary market could amount to millions of dollars, which potentially represents a loss in auction revenue. This could also encourage speculation in the auction, resulting in further rents and misallocations in future auctions.

So far we did not consider the possibility a speculator bidding in the auction. The following remark shows the role of a speculator when assignments are efficient.

\begin{remark}
    The speculator $s$ can never make a positive profit from bidding in an auction that assigns the units efficiently.
\end{remark}

Given that the speculator has zero use value for all the units, then they would automatically receive no unit in an efficient assignment.

\begin{proposition}\label{pro4}
The existence of speculation in the auction would increase the clearing price, ceteris paribus. 
\end{proposition}
\begin{proof}
As per our basic setup consider an auction where $n$ firms compete for $k$ homogeneous permits. Each firm $i$ has a private valuation $v_i$ for the permits, and submits a vector of bids (demand schedule) $\mathbf{b}_i$. Denote $\mathbf{b}^*$ as the equilibrium bids submitted by all bidders without the speculator. Suppose the speculative bidder $s$, enters the auction and bids $\mathbf{b}_s$. First note that other bidders do not have any incentives to reduce their bids compared to a case without the speculator. The bids submitted by speculator can either be non binding meaning that they are not among the top $k+1$ bids or binding, that is, at least one of them is among the top $k+1$ bids. In the first case neither of the bidders have incentives to change their bids. In the second case, if $\mathbf{b}^*_i$ was the best response of bidder $i$ without a speculator, then reducing bids cannot be a best response for bidder $i$ when adding the bids by the speculator. 

Next we show there exist cases in which the clearing price increases. First denote $\mathbf{b}^{**}$ as the new equilibrium vector of bids by all bidders including the speculator. It is clear that the first $k+1$ elements of $\mathbf{b}^{**}$ are at least as high as $\mathbf{b}^{*}$ and therefore the new auction clearing price is at least as high as before. If at least one of the bids by the speculator is among the top $k$ bids then the new clearing price is strictly higher than the previous one.

\end{proof}

The above results suggests that the existence of speculators in the auction would result in an upward shift of the auction clearing price. Note that since the speculator has zero use value for the units, this can only reduce the allocative efficiency of the auction. Of course, if the auction allocates the objects efficiently then speculators cannot win any unit in the auction and their existence would not change the outcome. Therefore, the result of Proposition \ref{pro4} is another support for the importance of the efficient initial allocation of permits.

\section{Conclusive remarks}

This paper emphasizes the necessity of designing auction mechanisms that encourage truthful bidding and efficient initial allocations. However, recognizing the limitations of any design, the establishment of a robust secondary market becomes imperative. Policymakers must ensure that these markets are transparent, liquid, and free from manipulation to effectively serve their role in reallocating permits efficiently.

Furthermore, understanding the interactions between primary auctions and secondary markets can guide the introduction of regulatory measures such as price floors or ceilings, market maker roles, or other forms of intervention intended to enhance market stability and efficiency.

The next logical step in this line of inquiry, as suggested, involves examining scenarios where the secondary market might compensate for an inefficient initial allocation. Future propositions and research could focus on quantifying the extent of efficiency restoration possible through secondary trading and identifying the conditions under which these markets most effectively enhance overall welfare. Additionally, exploring the impact of different types of secondary market structures (e.g., over-the-counter vs. exchange-based trading) on the efficacy of these corrective mechanisms could provide valuable insights for both academics and practitioners in environmental economics.
\newpage
%%%%%%%%%%%%%%%%%%%%%%%%%%%%
%%%%%%%%%%%%%%%%%%%%%%%%%%%%
\bibliographystyle{elsarticle-harv}
\bibliography{emission}

\end{document}